\documentclass[%
 reprint,
 amsmath,amssymb,
 aps,
]{revtex4-1}
\usepackage{url}
\usepackage[colorlinks,linkcolor=blue]{hyperref}
\usepackage{graphicx}
\usepackage{dcolumn}
\usepackage{bm}

\begin{document}

\title{Effective Rabi dynamics of Rydberg atoms and robust high-fidelity quantum gates with a resonant amplitude-modulation field}
\author{Jin-Lei Wu$^{1}$}\author{Shi-Lei Su$^{2}$}\author{Yan Wang$^{1}$}\author{Jie Song$^{1}$}\email[]{jsong@hit.edu.cn}\author{Yong-Yuan Jiang$^{1}$}\author{Yan Xia$^{3}$}
\affiliation{$^{1}$School of Physics, Harbin Institute of Technology, Harbin 150001, China\\
$^{2}$School of Physics, Zhengzhou University, Zhengzhou 450001, China\\
$^{3}$Department of Physics, Fuzhou University, Fuzhou 350002, China}
\begin{abstract}
With a resonant amplitude-modulation field on two Rydberg atoms, we propose a Rydberg antiblockade~(RAB) regime, where the Rabi oscillation between collective ground and excited states is induced. A controlled-Z gate can be yielded through a Rabi cycle. Further, several common issues of the RAB gates are solved by modifying the parameter relation. The gate fidelity and the gate robustness against the control error are enhanced with a shaped pulse. The requirement of control precision of the Rydberg-Rydberg interaction strength is relaxed. In addition, the atomic excitation is restrained and therefore the gate robustness against the atomic decay is enhanced.
\end{abstract}
\maketitle

\section{Introduction}\label{Sec1}
Since the pioneering proposal of performing quantum gates in a Rydberg atom system was reported~\cite{PhysRevLett.85.2208}, quantum computation with Rydberg atoms has been attractive and promising due to the long lifetimes of internal atomic states~\cite{Gallagher1994,Saffman2010}. Due to the powerful Rydberg-Rydberg interaction~(RRI), at most one atom within the blockade distance can be excited into Rydberg states, which is termed ``Rydberg blockade". A Rydberg superatom can be formed and the quantum information processing in a mesoscopic scale is potential by storing information in collective states of mesoscopic ensembles~\cite{PhysRevLett.87.037901}.

Beyond the Rydberg blockade, the Rydberg antiblockade~(RAB) has also great potential applications in quantum physics, such as gaining the quantitative strength information of the RRI~\cite{Ates2007}, carrying out quantum computation~\cite{PhysRevLett.85.2208,Su201702}, and creating steady entangled states~\cite{Carr2013,Su2015,Shao2017}. There are usually two ways to the complete RAB where the double excitation state $|rr\rangle$ can be fully populated. One is to use a strong external field that drives the excitation of Rydberg atoms with a Rabi frequency far larger than the RRI strength $V$~\cite{PhysRevLett.85.2208}. The other is based on a strong RRI much stronger than the Rabi frequency, but a large detuning $\Delta$ compensates for the RRI~\cite{Zuo2010,Lee2012,Li2013,Carr2013,Su2016,Su2017,Zhu2019,wu2019,Shao2017,Li:19}. The latter can be used to construct one-step quantum logic gates~\cite{Su2016,Su2017}, but it suffers from several obstacles of implementing high-fidelity quantum gates. Firstly,the required parameter relation is so strict that even a minuscule error in some parameters such as the RRI strength or detuning is unbearable~\cite{Su2016,Su2017,Zhu2019,wu2019,Shao2017,Li:19}. Secondly, when atoms are doubly excited the gradient of the RRI potential will cause a strong interatomic force that can induce the mechanical motion and decoherence of the atoms~\cite{Li2013}, which inevitably reduces the gate fidelity. Finally, during the gate operation, the population of Rydberg states may lead to decoherence~\cite{PhysRevA.85.042310}. Especially for the case where the Rydberg states are shelved for a while~\cite{PhysRevLett.85.2208,PhysRevA.98.052324,PhysRevA.97.032701,Shen:19,Zhu2019,Yu:19,Liao:19}, the decoherence effect is significant.

In this work, we propose to realize the complete RAB with a resonant amplitude-modulation field. Different from the conventional ways, the external field motivates the atomic excitation resonantly with a Rabi frequency much less than the RRI strength. This complete RAB yields the Rabi oscillation between the collective ground and excited states of the two atoms. For applications, it can be employed in creating steady entangled states with the assistance of a microwave field~\cite{Su2015}. Alternatively, we here mainly focus on a controlled-Z~(CZ) gate formed straightforwardly through a single Rabi cycle. Similar to the existing RAB schemes~\cite{Su2016,Su2017,Zhu2019,wu2019,Shao2017,Li:19,PhysRevLett.85.2208}, the implementation of this CZ gate may also encounter with the obstacles of the RAB gates.

In order to avoid the obstacles of the RAB gates, we propose to break the RAB by modifying the parameter relation between the RRI strength and the amplitude modulation frequency. The resulting dynamics is only dominated by a Stark shift of $|11\rangle$ that can induce arbitrary phase gates. Besides, the maximum amplitude of the Rabi frequency can be shaped so as to further strengthen the fidelity and robustness of the gates. To sum up, the present work has the following features:
(a)~The Rabi oscillation between $|11\rangle$ and $|rr\rangle$, i.e., the complete RAB, is achieved;
(b)~Arbitrary phase gates can be obtained by directly accumulating a phase on $|11\rangle$ without the attendance of Rydberg states;
(c)~The modified parameter relation allows for the pulse shaping so as to improve the fidelity and robustness of the gates;
(d)~The gate robustness against the atomic decay and the RRI strength error is enhanced since the population of $|rr\rangle$ is eliminated.

\section{Effective Rabi dynamics and controlled-Z gate}\label{Sec2}
\begin{figure}\centering
\includegraphics[width=\linewidth]{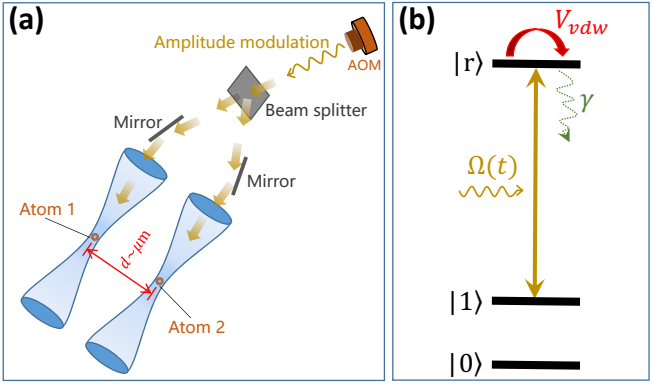}
\caption{(a)~Schematic diagram of two trapped Rydberg atoms irradiated synchronously with an amplitude-modulation external field. (b)~Schematic diagram of atomic levels and interaction.}\label{f1}
\end{figure}
As shown in Fig.~\ref{f1}(a), two identical Rydberg atoms apart from each other with the distance $d$ are confined in optical tweezers. We consider the van der Waals interaction~(vdWI) with a strength $V_{vdw}=C_6/d^6$. $C_6$ is the vdWI coefficient depending on the quantum numbers of the
Rydberg states~\cite{Gallagher1994,Saffman2010}. An amplitude-modulation field is imposed synchronously on the two atoms to drive the transition $|1\rangle\leftrightarrow|r\rangle$ with a Rabi frequency $\Omega(t)$, as shown in Fig.~\ref{f1}(b). Then the Hamiltonian of the two-atom system in the interaction picture can be represented by~($\hbar=1$ hereinafter)
\begin{eqnarray}\label{e1}
\hat H_I=\Big[\sum_{j=1,2}\frac{\Omega(t)}2|1\rangle_j\langle r|+{\rm H.c.}\Big]+V_{vdw}|rr\rangle\langle rr|.
\end{eqnarray}
The dynamics obeys the Schr\"{o}dinger equation $i\partial|\Psi(t)\rangle/\partial t=\hat H_I|\Psi(t)\rangle$.
Here we introduce the modulated Rabi frequency $\Omega(t)=\Omega_{0}\cos(\omega t)$ with $\Omega_{0}$ being the maximum amplitude and $\omega$ the modulation frequency. Such an amplitude-modulation field can be realized by an acousto-optic modulator~(AOM) with the help of the arbitrary waveform generator~\cite{dugan1997high}.

When the initial state $|11\rangle$ is considered, the two-atom Hamiltonian is
\begin{eqnarray}\label{e3}
\hat H_{11}&=&\frac{\Omega_{0}}{2}\cos(\omega t)\Big[|11\rangle(\langle1r|+\langle r1|)+(|1r\rangle+|r1\rangle)\langle rr|\nonumber\\
&&+{\rm H.c.}\Big]+V_{vdw}|rr\rangle\langle rr|.
\end{eqnarray}
After performing a unitary transformation $\hat {\mathcal{U}}_0\equiv\exp(i\hat h_0t)$ with $\hat h_0\equiv\omega_0|rr\rangle\langle rr|$, the Hamiltonian (\ref{e3}) becomes~\cite{PhysRevLett.120.123204}
\begin{eqnarray}\label{e4}
\hat {\mathcal{H}}_{11}
&=&\left\{\frac{\Omega_0}{2\sqrt2}\left(e^{i\omega t}+e^{-i\omega t}\right)|11\rangle\langle \Psi_{m}|+\frac{\Omega_0}{2\sqrt2}\big[e^{i(\omega-\omega_0) t}\right.\nonumber\\
&&\left.+e^{-i(\omega+\omega_0) t}\big]|\Psi_{m}\rangle\langle rr|+{\rm H.c.}\right\}\nonumber\\&&+(V_{vdw}-\omega_0)|rr\rangle\langle rr|,
\end{eqnarray}
in which $|\Psi_{m}\rangle\equiv(|r1\rangle+|1r\rangle)/\sqrt2$ is the intermediate single-excitation state. Intermediated by $|\Psi_{m}\rangle$, $|11\rangle$ is coupled to $|rr\rangle$ through four paths with detunings $\omega_0$, $\omega_0+2\omega$, $\omega_0-2\omega$, and $\omega_0$, respectively. In order to get the resonant $|11\rangle\leftrightarrow|rr\rangle$ interaction, $\omega_0=\pm2\omega$ should be chosen. By considering the second-order perturbation theory~\cite{James2007CJP} under the condition ${2\sqrt2}\omega\gg\Omega_0$, with $\omega_0=2\omega$ an effective Hamiltonian is developed
\begin{eqnarray}\label{e5}
\hat {\mathcal{H}}_{\rm eff}=[({\Omega_{\rm eff}}/{2})|11\rangle\langle rr|+{\rm H.c.}]
+V'|rr\rangle\langle rr|,
\end{eqnarray}
with $\Omega_{\rm eff}\equiv{\Omega_0^2}/{4\omega}$ and $V'\equiv V_{vdw}-2\omega+{\Omega_0^2}/{6\omega}$.
The Hamiltonian (\ref{e5}) is reduced into a further effective Hamiltonian $\hat {{H}}_{\rm e}=\frac{\Omega_{\rm eff}}{2}|11\rangle\langle rr|+{\rm H.c.}$ by setting $V_{vdw}=2\omega-{\Omega_0^2}/{6\omega}$. This effective Hamiltonian holds the Rabi dynamics between $|11\rangle$ and $|rr\rangle$, exhibiting the resonant-interaction-induced RAB within the condition $V_{vdw}\gg|\Omega(t)|$ that is different from the conventional RAB achieved with a largely detuned or strongly driven interaction. As an application, this complete RAB can be exploited for creating a steady entangled state $(|10\rangle-|01\rangle)/\sqrt2$ by introducing a weak microwave field driving the ground state transition $|0\rangle\leftrightarrow|1\rangle$, similar to Ref.~\cite{Su2015}. In the following, alternatively we mainly focus on its application in constructing logic gates.

\begin{figure}[htb]\centering
\includegraphics[width=\linewidth]{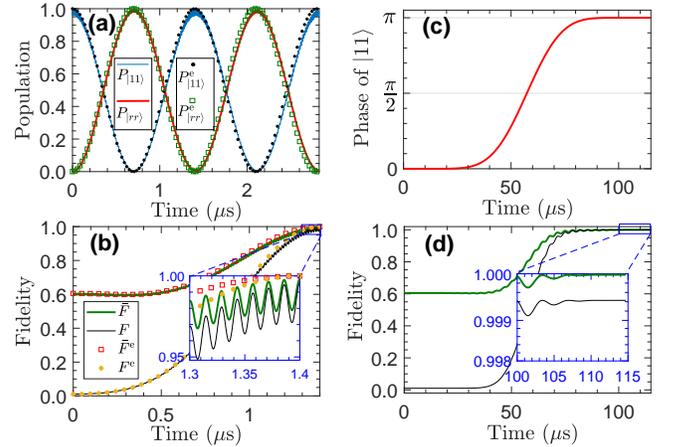}
\caption{Comparison between the full dynamics and the effective dynamics. The superscript ``${\rm e}$" denotes the effective dynamics. (a)~Rabi oscillations between the populations of $|11\rangle$ and $|rr\rangle$. (b)~Average and initial-state-specified fidelity of the CZ gate. (c)~Time evolution of the phase of $|11\rangle$. (d)~Average and initial-state-specified fidelity of the CZ gate. Parameters: $\Omega_0=2\pi\times10$ MHz, $\omega=2\pi\times35$ MHz, and $\delta=2\pi\times8$ MHz.}\label{f2}
\end{figure}
To illustrate the validity of the effective Hamiltonian (\ref{e5}), in Figs.~\ref{f2}(a) and (b) we simulate the comparison between the full dynamics and the effective dynamics, corresponding to Eqs.~(\ref{e1}) and (\ref{e5}), respectively. The Rabi dynamics between $|11\rangle$ and $|rr\rangle$ is shown in Fig.~\ref{f2}(a). The population evolution obtained by Eq.~(\ref{e1}) is almost coincident with the effective one, which proves the validity of the effective Hamiltonian.

A $\pi$ phase on $|11\rangle$ can be attained after a single Rabi cycle at the time $T_0=2\pi/\Omega_{\rm eff}$. When at most one atom is prepared initially in $|1\rangle$, the two-atom interaction is absent during the whole process. Governed by the evolution operator $\hat U=\hat{\mathcal{T}}\exp[-i\int_0^t\hat H_0(t')dt']$ with the single-atom Hamiltonian $\hat H_0(t)=\frac{\Omega(t)}2|1\rangle\langle r|+{\rm H.c.}$ and the time-order operator $\hat{\mathcal{T}}$, the evolutive state of the atom prepared initially in $|1\rangle$ can be expressed by $|\psi(t)\rangle=\cos\theta(t)|1\rangle-i\sin\theta(t)|r\rangle$ with $\theta(t)\equiv\sin(\omega t){\Omega_0}/{2\omega}$. Under the condition $2\omega\gg\Omega_0$~($|\theta(t)|\simeq0$), $|\psi(t)\rangle$ gets trapped in $|1\rangle$. As for the atom prepared initially in $|0\rangle$, its state will never evolve. Therefore, a CZ gate is induced at the time $t=T_0$,
\begin{eqnarray}\label{e8}
\hat U_{\rm CZ}=|00\rangle\langle00|+|01\rangle\langle01|+|10\rangle\langle10|-|11\rangle\langle11|.
\end{eqnarray}

As a test, we plot the average fidelity of the CZ gate in Fig.~\ref{f2}(b), for which we define the average fidelity as
\begin{eqnarray}\label{e9}
\bar F=\frac1{(2\pi)^2}\int_{-\pi}^\pi\int_{-\pi}^\pi|\langle\Psi_i|\Psi(t)\rangle|^2 d\alpha_1d\alpha_2,
\end{eqnarray}
where $|\Psi_i\rangle\equiv\hat U_{\rm CZ}|\Psi(0)\rangle$ is the ideal state after the CZ gate on an initial state
$|\Psi(0)\rangle=\cos\alpha_1\cos\alpha_2|00\rangle+\cos\alpha_1\sin\alpha_2|01\rangle+\sin\alpha_1\cos\alpha_2|10\rangle+\sin\alpha_1\sin\alpha_2|11\rangle$.
This gate fidelity means the average of fidelities at each moment for an infinite number of input states with $\alpha_1$ and $\alpha_2$ distributed over $[-\pi,~\pi)$, which can demonstrate the randomicity of the input state.
The average fidelity increases with time, reaching $\bar F>0.995$ finally. Alternatively, we specify an initial state $|\Psi'\rangle=0.5|00\rangle+0.5|01\rangle+\sqrt{0.05}|10\rangle+\sqrt{0.45}|11\rangle$ so as to make the gate fidelity $F=|\langle\Psi_i|\Psi(t)\rangle|^2$ close to zero at $t=0$, and the gate fidelity can also reach over 0.995 finally.

This RAB gate is different from the typical three-step Rydberg-blockade gate~\cite{PhysRevLett.85.2208}. Firstly, the gate is performed by just one step. Then, it does not need the individual laser addressing of the two atoms. In addition, the two atoms are fully excited just transiently. But for the three-step Rydberg-blockade gate, after the first step the control atom must be fully excited for a while to wait the end of the second step when it is in $|1\rangle$ initially, which may cause more decoherence.
However, this RAB gate also suffers several shortcomings:
(i)~The final fidelity is not high enough;
(ii)~The fidelity increases with significant oscillations, so it is sensitive to the control errors in operation time;
(iii)~The two atoms must be excited during the evolution, which causes decoherence due to atomic decay;
(iv)~The condition $V_{vdw}=2\omega-{\Omega_0^2}/{6\omega}$ is so strict that the CZ gate can not tolerate even a minuscule error~\cite{Su2016,Su2017,Zhu2019};
(v)~Once $|rr\rangle$ is occupied, the strong interatomic force may induce the mechanical motion and decoherence of atoms~\cite{Li2013}.

\begin{figure}[htb]\centering
\includegraphics[width=\linewidth]{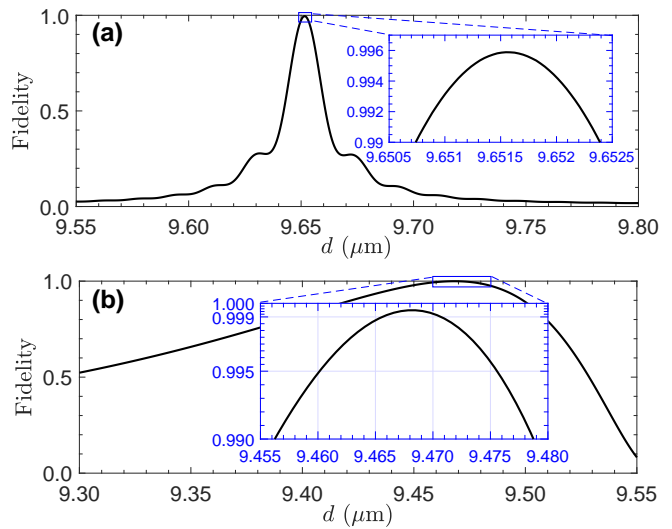}
\caption{Fidelity of the CZ gate with different distance between two atoms for (a)~the RAB and (b)~breaking the RAB. Parameters: $\Omega_0=2\pi\times10$ MHz, $\omega=2\pi\times35$ MHz, $\delta=2\pi\times8$ MHz, and $C_6=2\pi\times56.2~{\rm THz}~\mu {\rm m}^6$.}\label{f3}
\end{figure}
The points (i) and (ii) can be clearly found in Fig.~\ref{f2}(b). (iii) will be illustrated later. As for (iv) and (v), we assume that $^{87}$Rb atoms are adopted and the related energy levels are chosen as $|0\rangle=|5S_{1/2}, F = 1, m_{F}=0\rangle$, $|1\rangle=|5S_{1/2}, F = 2, m_{F}=0\rangle$, and $|r\rangle=|100S_{1/2}, m_{J}=1/2, m_{I}=3/2\rangle$~\cite{PhysRevApplied.11.044035}. For such Rydberg atoms, the vdWI coefficient is $C_6=2\pi\times56.2~{\rm THz}~\mu {\rm m}^6$~\cite{PhysRevA.77.032723,PhysRevA.90.062327}. Then in Fig.~\ref{f3}~(a), we show the sensitivity of the CZ gate to the interatomic distance, for which $|\Psi'\rangle$ is chosen as the initial state. The deviation range of $d$ keeping $F>0.99$ is less than $2$ nm. In fact, the extreme sensitivity to errors in the interatomic distance as well as the RRI strength is a common issue for the RAB gate schemes~\cite{Su2016,Su2017,Zhu2019,wu2019}. The shortcomings itemized above of implementing the CZ gate can be overcome by breaking the RAB, as shown in the next section.

\section{Arbitrary phase gates by breaking the Rydberg antiblockade}\label{Sec3}
For solving the problems (iii)$\sim$(v), one way is to suppress the atomic excitation by eliminating $|rr\rangle$ population~\cite{PhysRevLett.85.2392}. We modify the parameter condition $V_{vdw}=2\omega-{\Omega_0^2}/{6\omega}$ into $V_{vdw}-2\omega=\delta$ and $|\delta|\gg|\Omega_{\rm eff}|$. Then, because the Stark shift $|{\Omega_0^2}/{6\omega}|$ of $|rr\rangle$ is far smaller than $|\delta|$, it can be dropped safely. The largely detuned $|11\rangle\leftrightarrow|rr\rangle$ transition will induce a final effective Hamiltonian
\begin{eqnarray}\label{e10}
\hat {{H}}_{\rm f}=-({\Omega_{\rm eff}^2}/{4\delta})|11\rangle\langle 11|,
\end{eqnarray}
which solely involves a high-order Stark shift of $|11\rangle$. Since the $\Omega_0$-dependence of $V_{vdw}$ is avoided, the pulse shaping can be performed for $\Omega_0$ to solve the problems (i) and (ii). The waveform of a single-period cos-like function is applicable
\begin{eqnarray}\label{e12}
\Omega_0(t)={\Omega_{\rm m}}[1-\cos({2\pi t}/{T})]/2,
\end{eqnarray}
with $\Omega_{\rm m}$ being the maximum and $T$ the period~(gate duration). Then governed by $\hat {{H}}_{\rm f}$, two-qubit arbitrary phase gates can be attained
\begin{eqnarray}\label{e11}
\hat U_{\rm phase}=|00\rangle\langle00|+|01\rangle\langle01|+|10\rangle\langle10|+e^{i\varphi}|11\rangle\langle11|,
\end{eqnarray}
with $\varphi={\int\Omega_{\rm eff}^2dt}/{4\delta}$.

As an example, we conduct a CZ gate with $\varphi=\pi$ and $T=234\pi\omega^2\delta/\Omega_{\rm m}^4$. Based on the full Hamiltonian Eq.~(\ref{e1}), we plot the phase change of the state $|11\rangle$ in Fig.~\ref{f2}(c), in which the phase of $|11\rangle$ is stabilized at $\pi$ finally. Correspondingly, the average fidelity and the initial-state-specified~($|\Psi'\rangle$) fidelity are given in Fig.~\ref{f2}(d). We can clearly see that at the cost of a longer gate time the final average fidelity is stationary and reaches over $0.9999$.
In Fig.~\ref{f3}(b) we plot the sensitivity of the CZ gate to the interatomic distance with the specified initial state $|\Psi'\rangle$, for which we consider the same candidate atoms and energy levels as those for Fig.~\ref{f3}(a). In Fig.~\ref{f3}(b) the deviation range of $d$ keeping $F>0.99$ is more than $20$ nm and even $F>0.999$ about $5$ nm. In Fig.~\ref{f3}, the adopted parameters indicate $V_{vdw}\sim100\Omega_{\rm eff}$ and $V_{vdw}\sim10\delta$. For the RAB CZ gate, $1\%$ deviation in $V_{vdw}$ results in the off-resonant $|11\rangle\leftrightarrow|rr\rangle$ transition with the detuning being of the same order as $\Omega_{\rm eff}$, which greatly spoils the complete Rabi cycles~\cite{PhysRevApplied.11.044035} and then reduces significantly the gate fidelity by over 0.5. However for the RAB-broken gate, $1\%$ deviation in $V_{vdw}$ results in $\sim10\%$ deviation in $\delta$. This deviation changes the phase of $|11\rangle$ by just $\sim0.01$ and thus reduces seldom the gate fidelity. Therefore, the RAB-broken CZ gate displays better robustness against the deviation in the RRI than the RAB CZ gate.

\begin{figure}[htb]\centering
\includegraphics[width=\linewidth]{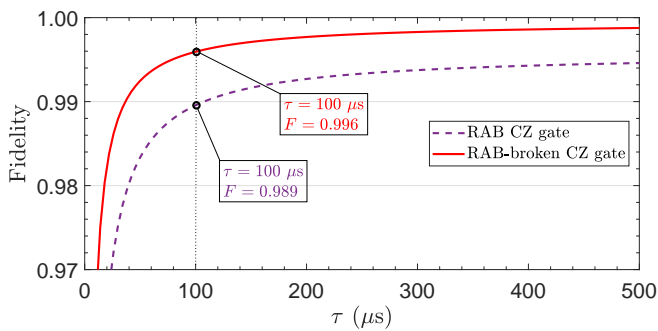}
\caption{Influence of the Rydberg state lifetime on the final fidelity of the RAB CZ gate~(dashed line) and the RAB-broken CZ gate~(dashed line).}\label{f4}
\end{figure}
Taking the decay of the Rydberg states into account, the evolution is dominated by the master equation
\begin{eqnarray}\label{e13}
{\partial{\hat{\rho}}}/{\partial t}&=&i\hat{\rho}\hat{H}_I-i\hat{H}_I\hat{\rho}\nonumber\\
&&-\frac{1}{2}\sum^{4}_{k=1}\left(\hat{\mathcal{L}}_{k}^{\dag}\hat{\mathcal{L}}_{k}\hat{\rho}-2\hat{\mathcal{L}}_{k}\hat{\rho}\hat{\mathcal{L}}_{k}^{\dag}
+\hat{\rho}\hat{\mathcal{L}}_{k}^{\dag}\hat{\mathcal{L}}_{k}\right),
\end{eqnarray}
where $\hat{\rho}$ is the density operator. $\hat{\mathcal{L}}_{1}\equiv\sqrt{\gamma/2}|0\rangle_{1}\langle r|$, $\hat{\mathcal{L}}_{2}\equiv\sqrt{\gamma/2}|1\rangle_{1}\langle r|$, $\hat{\mathcal{L}}_{3}\equiv\sqrt{\gamma/2}|0\rangle_{2}\langle r|$, and $\hat{\mathcal{L}}_{4}\equiv\sqrt{\gamma/2}|1\rangle_{2}\langle r|$ are Lindblad operators describing four decay paths. $\gamma\equiv1/\tau$ is the atomic decay rate with $\tau$ being the lifetime of the Rydberg state. Here we suppose that the decay rates from $|r\rangle$ to $|0\rangle$ and $|1\rangle$ for each atom
are $\gamma/2$ identical. In Fig.~\ref{f4} we plot the influence of the Rydberg state lifetime on the final fidelities of the RAB CZ gate and the RAB-broken CZ gate, for which the initial state $|\Psi'\rangle$ is adopted and the fidelity is defined by $F(t)=\langle\Psi'|U_{\rm CZ}^\dagger\hat{\rho}U_{\rm CZ}|\Psi'\rangle$. For the error correction threshold $F=0.99$ in surface code schemes~\cite{PhysRevA.80.052312}, $\tau=40~\mu$s is enough for the RAB-broken CZ gate. For an accessible Rydberg state lifetime $\tau=100~\mu$s~\cite{PhysRevA.79.052504}, the final fidelity of the RAB-broken CZ gate reaches $0.996$. As for the RAB CZ gate, even with $\tau=100~\mu$s the final fidelity is less than $0.990$.
Therefore, the gate robustness against the atomic decay is strengthened by breaking the RAB.

\section{Conclusion}\label{Sec4}
The effective Rabi dynamics between $|11\rangle$ and $|rr\rangle$ of the two Rydberg atoms is proposed by using a resonant amplitude-modulation field. The induced complete Rydberg antiblockade is different from the conventional ones. By breaking the Rydberg antiblockade, two-qubit arbitrary phase gates can be obtained. By means of the shaped pulse, the arbitrary phase gates are of not only high fidelity but also great control error tolerance at the cost of a longer gate time. Besides, the breakdown of the Rydberg antiblockade improves the gate resilience against the deviation of the interatomic distance, and enhances the gate robustness to the atomic decay. Finally, we hope our work could be helpful for the experimental researches in Rydberg excitation and quantum computing.

\section*{ACKNOWLEDGEMENTS}
This work was supported by National Natural Science Foundation of China (NSFC) (11675046), Program for Innovation Research of Science in Harbin Institute of Technology (A201412), and Postdoctoral Scientific Research Developmental Fund of Heilongjiang Province (LBH-Q15060).

\bibliography{apssamp}
\end{document}